# Magnetic Field of Firing Neuron in Humans: Measurable by Quantum Sensing MRI?

Yongxian Qian [1, 2, 3, *] OCRID: 0000-0002-9931-7790

[1] Bernard and Irene Schwartz Center for Biomedical Imaging, Department of Radiology;

[2] Center for Advanced Imaging Innovation and Research (CAI$^2$R), Department of Radiology;

[3] Neuroscience Institute, Department of Neuroscience and Physiology;

New York University Grossman School of Medicine, New York, New York 10016, USA.

* **Corresponding author**. Email: Yongxian.Qian@nyulangone.org

**Conflict of interest:** YQ is one of the inventors of U.S. Patent Application: No.: 63/467,482. *Systems and methods for non-invasive detection of neuronal firings in humans via quantum-sensing magnetic resonance imaging*. Filed on May 18, 2023. IP Owner: New York University.

**Funding:** National Institutes of Health (NIH) RF1/R01 AG067502; the General Research Fund (GRF) from the Department of Radiology, NYU Grossman School of Medicine.

**Acknowledgments:** The author thanks the following colleagues and peers for their insightful discussions about neuronal magnetic fields and MRI detectability: Dr. Tom Barbara at the Advanced Imaging Research Center, Oregon Health & Science University, Portland, OR; and Dr. Ravinder Regatte at the Department of Radiology, NYU Grossman School of Medicine, New York, NY. The author is grateful to the anonymous reviewers for their critical but inspiring comments on the quantum sensing MRI proposed in our paper manuscripts and grant applications. This work was conducted under the rubric of the Center for Advanced Imaging Innovation and Research (CAI2R), a National Institute of Biomedical Imaging and Bioengineering (NIBIB) Biomedical Technology Resource Center grant NIH P41 EB017183.

**Abstract:** Firing neurons generate action potentials that propagate along axons to transmit signals supporting cognitive functions. These electrical currents generate magnetic fields, yet direct detection of these neuronal magnetic fields by MRI remains elusive. This Mini Review investigates why this goal has been proven difficult to achieve and whether an emerging approach – quantum sensing MRI – can overcome the challenge.

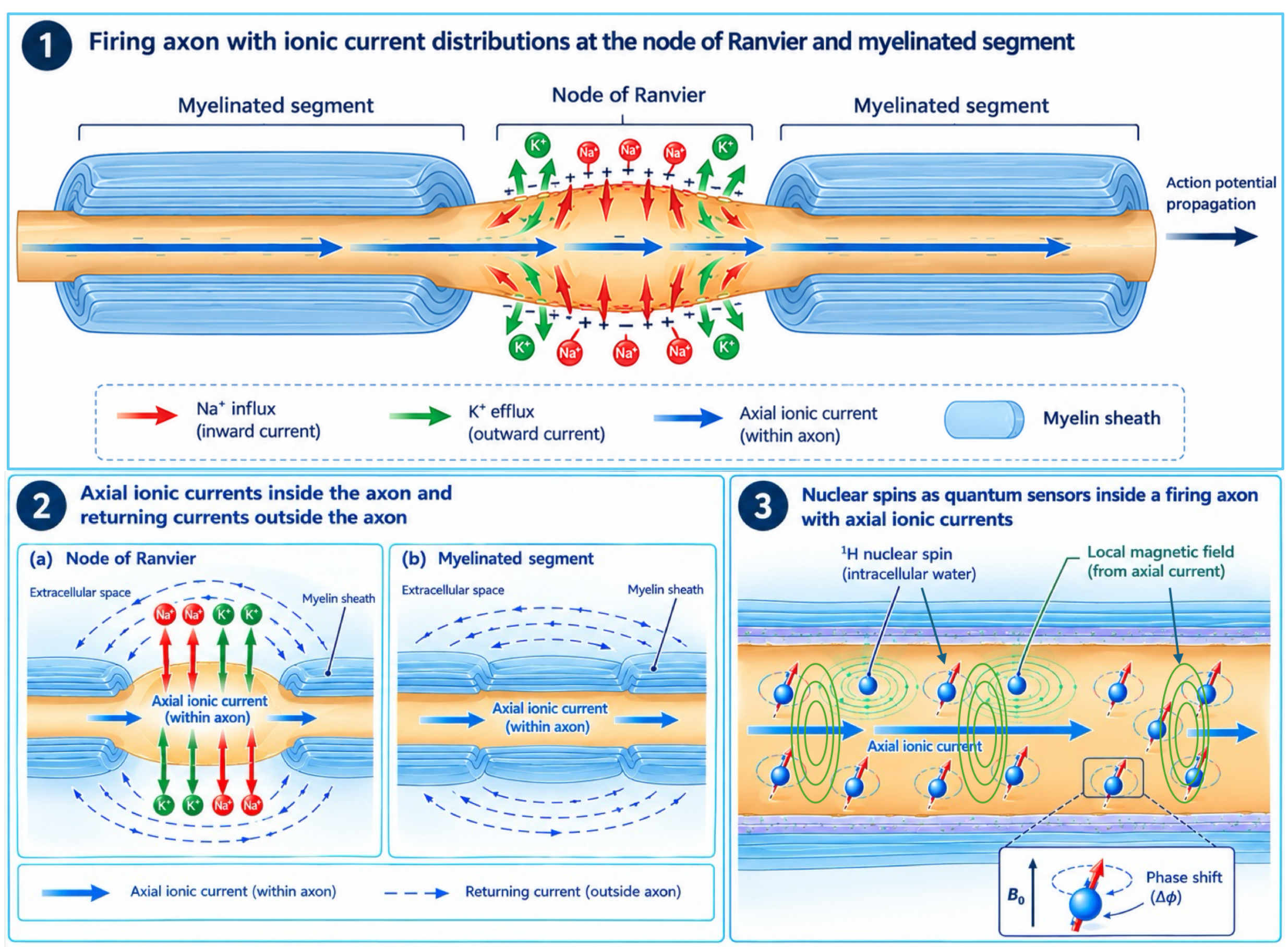

**Figure 1. Firing axon and its ionic currents. Panel 1:** Ionic influx and efflux at successive nodes of Ranvier regenerate the action potential as it propagates along the myelinated axon. **Panel 2:** Ionic current distributions at the node of Ranvier (**a**) and along a myelinated segment (**b**), showing the corresponding extracellular return currents. **Panel 3:** Intracellular water proton ($^1H$) nuclear spins serving as quantum sensors for detecting local magnetic fields generated by axial ionic currents.

## Firing neurons and associated magnetic fields

Neuronal firing is the rapid electrical activity that generates action potentials initially at the axon hillock then conducting along the axon to transmit information between neurons and ultimately support brain functions such as perception, memory, language, and cognition. An action potential is generated by coordinated transmembrane ionic currents (membrane currents), dominated by sodium ion ($Na^+$) influx during depolarization and potassium ion ($K^+$) efflux during repolarization (Figure 1, Panel 1), and typically lasts approximately 1–2 ms.[1–4] Because moving electric charge generates a magnetic field, neuronal firing necessarily produces a transient magnetic field that propagates with the action potential.

The magnetic field generated by a firing neuron is determined by the spatial distribution and temporal dynamics of both transmembrane ionic currents and intracellular axial ionic currents (axial currents). For an idealized cylindrical axon carrying an axial current ($I_a$), the azimuthal magnetic field outside the axon is $B=\mu_0 I_a/(2\pi r)$, where $r$ is the distance from the axonal axis.[5,6] Real neurons are considerably more complex, containing branching dendrites, axons, nodes of Ranvier, myelin, and distributed return currents. Nevertheless, the fundamental principle remains: neuronal electrical activity is inevitably accompanied by magnetic fields.

## Neuronal magnetic field outside an axon

This topic has been extensively investigated through both computational modeling and experimental measurements. Outside an axon, the magnetic field associated with an action potential arises primarily from transmembrane and axial ionic currents. The extracellular field decreases rapidly with distance from the current sources and can undergo substantial spatial cancellation because ionic currents flowing across different portion of the membrane have opposing directions and spatially varying orientations.[5–9] Computational models of single axons and nerve fibers have demonstrated that extracellular magnetic waveforms can be derived from the underlying transmembrane action potential.[7,10] Magnetoencephalography (MEG) detects neuronal magnetic fields at the scalp, typically at sub-picotesla levels (~0.2 pT), with millisecond temporal resolution.[11,12] These magnetic fields arise from synchronized neuronal currents. At the level of individual axons or nerve fibers, magnetic fields have been measured experimentally at approximately 0.06 nT near the frog sciatic nerve [13] and modeled to reach approximately 0.6 nT around a large axon with a diameter of 200 μm.[14] These measurements demonstrated that action potentials – and the associated transmembrane currents – generate magnetic fields outside the membrane at sub-nanotesla to nanotesla levels (approximately 0.06–20 nT). Such weak fields produce only minute phase shifts in the proton ($^1$H) MRI signal; for example, a 2-ms sampling interval ($\Delta t$) would yield a phase shift of approximately 0.002º–0.60º at these field strengths, according to $\Delta\varphi = \gamma B_n \Delta t$ and gyromagnetic ratio $\gamma$=42.58 MHz/T. This extreme weakness of the extracellular field represents a fundamental barrier to direct MRI detection of neuronal activity.

Despite the weakness of these fields, MRI researchers have attempted to detect neuronal magnetic fields directly. Early theoretical studies showed that neuronal currents can produce phase shifts in MR signal and predicted that phase-based detection could offer greater sensitivity than magnitude-based detection. Petridou et al. demonstrated detection of transient magnetic fields as small as approximately 0.2 nT using a current wire phantom,[15] while Konn et al. demonstrated MR phase changes induced by an extended current dipole,[16] providing experimental support for the feasibility of phase-based MRI detection of weak magnetic fields. Human studies subsequently reported possible MRI signals associated with spontaneous alpha activity,[17] but independent investigations failed to reproduce these findings.[18,19] Other studies likewise reported null findings or substantial uncertainty,[20,21] underscoring the limited sensitivity of conventional MRI for direct detection of neuronal currents.

The difficulty arises from several factors, including the extremely weak extracellular fields, spatial averaging, contribution of opposing phase distributions within a voxel, the orientation of neuronal currents relative to the main magnetic field $\boldsymbol{B}_0$, physiological noise, and the mismatch between millisecond-scale neuronal events and the tens- to hundreds-of-milliseconds temporal resolution of conventional MRI. Thus, despite extensive investigation, direct MRI detection of extracellular neuronal magnetic fields has not been reproducibly validated in humans.

## Neuronal magnetic field inside an axon

The intra-axonal magnetic field has been much less studied than the extra-axonal field because direct experimental measurement inside a firing axon remains technically challenging, whether invasively or noninvasively. One important challenge is the surrounding return current: ionic currents flowing outside the axon can partially cancel the magnetic field generated by the intracellular axial current when measured externally (Fig. 1, Panel 2).[1,13] Consequently, measurements outside the axon may substantially underestimate the local magnetic field experienced by nuclear spins inside the axon. Nevertheless, its existence is firmly established by electromagnetic theory: the ionic currents associated with an action potential – including both transmembrane and axial currents – necessarily generate magnetic fields. The key unresolved question is therefore *how large this intra-axonal filed is* – at the nano- or microtesla scale. These vastly different field strengths have major implications for MRI detectability and warrant rigorous theoretical and experimental investigation.

Computational modelling of a firing axon using Maxwell's equations, constrained by a prescribed action potential waveform, has estimated intra-axonal magnetic fields at sub-nanotesla to nanotesla levels,[8,10,14] comparable to those measured or modeled outside the axon. These estimates suggest that the intra-axonal fields may be too week for detection with current MRI approaches.[16-21] These nanotesla estimates may, at least in part, reflect the constraint imposed by the action potential, which is conventionally defined by the voltage *across the membrane* and therefore primarily constraints transmembrane ionic currents. Such a formulation may not fully capture the contribution of axial ionic currents along the axon, which could generate substantially larger magnetic fields within the axoplasm.

The intra-axonal magnetic field generated by axial ionic currents can, however, be estimated from several characteristic parameters governing action potential conduction. First, axial currents arise from the influx of $Na^+$ ions at the node of Ranvier, with the resulting longitudinal current propagating along the axon and contributing to saltatory conduction of the action potential. Because membrane currents are distributed around the circumference of the axon, their longitudinal contribution can accumulate to produce an axial current substantially greater than the current crossing any individual membrane segment. For a cylindrical axon, the integrated contribution around the circumference introduces a geometric factor of approximately $2\pi$ relative to the corresponding current per unit azimuthal angle. Second, the magnitude of the axial current can be related to the longitudinal transport of ionic charge associated with action potential. For a propagating charge distribution, the axial current can be expressed approximately as $I_a = \lambda v$, where $\lambda$ is the longitudinal ionic charge density and $v$ is the propagation velocity. In myelinated axons, action potentials are regenerated at successive nodes of Ranvier and propagate between nodes through rapid longitudinal redistribution of intracellular charge, supported by both electrochemical (concentration) gradients and electrical forces established between the depolarized node and the downstream polarized node.[1,2] Because conduction velocities in myelinated axons can be ~100–300 times greater than those in unmyelinated axons,[1,2] the associated axial charge transport can likewise be substantially enhanced, provided that the longitudinal charge density is comparable. Thus, the axial current may be considerably greater than the current crossing an individual membrane segment, providing a potential mechanism for generating intra-axonal magnetic fields substantially larger than estimates based solely on transmembrane currents. Third, the geometry of current flow along a myelinated axon may further enhance the intra-axonal magnetic field. Unlike transmembrane currents, which are distributed radially across the axonal membrane and may partially cancel because of their opposing directions around the circumference, axial currents are predominantly longitudinal and therefore produce magnetic fields that can add constructively within the axon. The combination of a relatively large longitudinal charge transport current and its coherent spatial organization provides a plausible mechanism for generating an intra-axonal magnetic field substantially stronger than that predicted from membrane currents alone. Importantly, the magnitude of this field depends on axonal radius, current density, conduction velocity, and the spatial extent over which the axial currents remain coherently oriented; therefore, the proposed enhancement should be regarded as an order-of-magnitude physical estimate rather than a direct equivalence between conduction velocity and current amplitude.

These considerations provide a physical basis for estimating the magnitude of the magnetic field generated by a firing axon. For a long, approximately cylindrical myelinated axonal segment, the magnetic field produced by an axial current can be estimated from Ampère's law, with the field magnitude determined primarily by the axial current and the radial distance from the current. During action potential propagation, the axial current is concentrated within the axoplasm and extends over the relatively long internodal segment between adjacent nodes of Ranvier. Because an internodal segment can be ~200 μm long, compared with a node of Ranvier of ~1 μm, the longitudinal distribution of axial current provides a much larger spatial extent for the generation of an intra-axonal magnetic field than the localized transmembrane current at an individual node.

Using experimentally and physiologically plausible parameters for axonal geometry, ionic current, and action potential propagation, the resulting intra-axonal magnetic field can be estimated to be on the order of tens to hundreds of microtesla, such as 18–182 μT for the range of parameters considered. This estimate is substantially larger than the subnano- to nano-tesla fields predicted by models constrained primarily by the transmembrane action potential waveform. The difference may arise because the latter models primarily constrain the voltage across the membrane, whereas the present estimate explicitly considers the longitudinal axial current associated with propagation along the axon. Although the precise field magnitude will depend on axonal diameter, current distribution, conduction velocity, and the spatial geometry of the propagating action potential, the analysis suggests that intra-axonal magnetic fields may be considerably larger than previously estimated from transmembrane currents alone.

Our recent study, currently available as an *arXiv* preprint and not yet peer reviewed, provided an independent quantitative estimate of the intra-axonal magnetic field.[22] Using a simplified long-straight-wire model to approximate the myelinated segment of an axon, we estimated magnetic field strengths of approximately 18–182 μT for axonal radii of 0.1–1.0 μm.[22] The model was motivated in part by the large length ratio (~200:1) between a myelinated segment (~200 μm) and a node of Ranvier (~1 μm long), which suggests that axial currents may extend over a substantially greater distance than the localized transmembrane currents at the nodes.[1,2] Importantly, microtesla-scale magnetic field variations, within approximately ±50 μT, were also detected in our preliminary human studies using quantum sensing MRI, a recently proposed MRI-based approach.[22]

Although these findings require independent validation and peer-reviewed confirmation, the convergence between the theoretical estimate and preliminary human measurements provides initial evidence that intra-axonal magnetic field during neuronal firing may reach microtesla range.

## Quantum sensing MRI for measuring magnetic fields inside firing neurons in humans

Quantum sensing MRI (qsMRI) [22] applies the general concept of quantum sensing [23] as a new strategy for detecting local magnetic field variations in vivo, while leveraging clinical MRI scanners as a scalable hardware platform. In qsMRI, radiofrequency (RF) pulses are used to manipulate the quantum states of nuclear spins in human tissue at room temperature, and the resulting spin dynamics are encoded in and detected through the MR signal. Importantly, qsMRI uses endogenous nuclear spins – including the abundant proton ($^1$H) nuclear spins of tissue water – as distributed quantum sensors that can encode and report local magnetic field variations (Fig. 1, Panel 3). This approach therefore provides a potential means of detecting magnetic fields generated by neuronal electrical activity without exogenous contrast agents. A nuclear spin precesses at the Larmor frequency, $\omega_0 = \gamma B_0$, in the main static magnetic field $\boldsymbol{B}_0$ of the MRI system. A transient neuronal magnetic field $\boldsymbol{B}_n(t)$ perturbs the nuclear spin Zeeman energy levels – which are quantum energy levels and motivate the term "*quantum sensing*" in qsMRI – through its component $B_{n,z}(t)$ along the direction of $\boldsymbol{B}_0$. This perturbation alters the Larmor frequency, producing an additional frequency shift and corresponding phase accumulation during the sampling interval $\Delta t$, i.e., $\Delta\varphi = \gamma B_{n,z}\Delta t$. Because this phase evolution is encoded in the Free Induction Decay (FID) signal, $B_{n,z}(t)$ can be estimated from the phase difference measured over a sampling interval $\Delta t$ through $B_{n,z} = \Delta\varphi / \gamma\Delta t$.

Central to qsMRI is the acquisition of FID signals at very high temporal resolution, potentially on the microsecond scale, to resolve transient phase evolution associated with neuronal electrical activity.[22] Unlike blood-oxygenation-level-dependent (BOLD) functional MRI (fMRI),[24,25] which measures a delayed hemodynamic response to neuronal activity, qsMRI seeks a magnetic signature generated directly by neuronal electrical activity – namely, the intra-axonal magnetic field associated with an action potential. This approach is potentially enabled by water proton nuclear spins within the intra-axonal compartment, which can serve as endogenous quantum sensors without materially perturbing the underlying ionic currents. More importantly, these nuclear spins may experience the microtesla-scale magnetic fields predicted within a firing axon, producing measurable changes in FID phase, whereas water proton nuclear spins outside the axons experience substantially weaker, nanotesla-scale neuronal magnetic fields that may be insufficient to produce a detectable phase change over the relevant sampling interval.

Our recent qsMRI work has yielded promising results in static water phantoms without neuronal firings and human subjects at rest and during finger-tapping motor tasks.[22] The proposed measurement chain is:

> *neuronal firing → intracellular axial ionic current → local magnetic field → nuclear spin phase change → ultrafast FID acquisition → neuronal magnetic field estimation*

Whether this measurement chain can reproducibly and specifically capture magnetic fields generated within firing axons remains a critical experimental question. Although our study provides experimental details to facilitate open validation,[22] independent replication will require rigorous control of electronic and physiological noise, quantitative electromagnetic modeling, and comparison with established electrophysiological measures, including electroencephalography (EEG), magnetoencephalography (MEG), and, where feasible, intracranial recordings. If independently validated, qsMRI could establish a new functional MRI contrast mechanism based on neuronal electrical activity rather than its downstream vascular consequences, potentially combining the anatomical and spatial capabilities of MRI with substantially faster temporal information about neuronal activity.

**Summary**

The ionic currents – including transmembrane and axial currents – that generate and propagate action potentials necessarily produce magnetic fields. The magnitude and spatial distribution of these fields differ substantially between the extracellular and intracellular spaces. Extracellular neuronal magnetic fields have been theoretically modeled and experimentally measured; however, their nanotesla-scale magnitude and spatial dispersion make their direct detection by MRI extremely challenging. In contrast, the magnetic field generated by concentrated axial ionic currents within an axon may reach the microtesla range, potentially producing measurable phase perturbations in nearby nuclear spins that serve as endogenous quantum sensors. This physical possibility provides a basis for the emerging concept of quantum sensing MRI, which may enable direct detection of neuronal magnetic fields in the human brain. Rigorous experimental validation, quantitative electromagnetic modeling, and independent replication will be essential to establish the reliability, specificity, and biological origin of signals attributed to neuronal magnetic fields. If validated, qsMRI could provide a fundamentally new functional MRI contrast mechanism based on neuronal electrical activity rather than its downstream vascular consequences.

**REFERENCES**

1. Koester JD & Siegelbaum SA, Membrane Potential and the Passive Electrical Properties of the Neuron. Bean BP & Koester JD, Propagated Signaling: The Action Potential. *In* Kandel ER, et al (Eds). *Principles of Neural Science*. 6th Ed. McGraw-Hill; 2021. pp.190–235.

2. Debanne D, Campanac E, Bialowas A, Carlier E, et al. Axon Physiology. *Physiol Rev*. 2011; **91**: 555–602.

3. Hodgkin AL, Huxley AF. A quantitative description of membrane current and its application to conduction and excitation in nerve. *J Physiol*. 1952; **117**:500–544.

4. Bean BP. The action potential in mammalian central neurons. *Nat Rev Neurosci*. 2007; **8**:451–465.

5. Swinney KR, Wikswo JP Jr. A calculation of the magnetic field of a nerve action potential. *Biophys J*. 1980; **32**:719–731.

6. Scott AC. The electrophysics of a nerve fiber. *Rev Mod Phys*. 1975; **47**:487–533.

7. Roth BJ, Wikswo JP Jr. The magnetic field of a single axon: a comparison of theory and experiment. *Biophys J*. 1985; **48**:93–109.

8. Wikswo JP, Van Egeraat JM. Cellular magnetic fields: fundamental and applied measurements on nerve axons, peripheral nerve bundles, and skeletal muscle. *J Clin Neurophysiol.* 1991; **8**:170–88.

9. Roth BJ, Wikswo JP Jr. The electrical potential and the magnetic field of an axon in a nerve bundle. *Math Biosci*. 1985; **76**:1–36.

10. Barach JP, Roth BJ, Wikswo JP. Magnetic measurements of action currents in a single nerve axon: A core-conductor model. *IEEE Trans Biomed Eng*. 1985; **28**:136–40.

11. Hämäläinen M, Hari R, Ilmoniemi RJ, Knuutila J, Lounasmaa OV. Magnetoencephalography – theory, instrumentation, and applications to noninvasive studies of the working human brain. *Rev Mod Phys*. 1993; **65**:413–497.

12. Baillet S. Magnetoencephalography for brain electrophysiology and imaging. *Nat Neurosci*. 2017; **20**:327–339.

13. Wikswo JP, Barach JP, Freeman JA. Magnetic field of a nerve impulse: first measurements. *Science*. 1980; **208(4439)**:53-55.

14. Woosley JK, Roth BJ, Wikswo Jr JP. The magnetic field of a single axon: A volume conductor model. *Math Biosci*. 1985; **76**:1-36.

15. Petridou N, Plenz D, Silva AC, Loew M, Bodurka J, Bandettini PA. Direct magnetic resonance detection of neuronal electrical activity. *Proc Natl Acad Sci USA*. 2006; **103**:16015–16020.

16. Konn D, Gowland P, Bowtell R. MRI detection of weak magnetic fields due to an extended current dipole in a conducting sphere: a model for direct detection of neuronal currents in the brain. *Magn Reson Med*. 2003; **50**:40–49.

17. Konn D, Leach S, Gowland P, Bowtell R. Initial attempts at directly detecting alpha wave activity in the brain using MRI. *Magn Reson Imaging*. 2004; **22**:1413–1427.

18. Parkes LM. Inability to directly detect magnetic field changes associated with neuronal activity. *Magn Reson Med*. 2007; **57**:411–416.

19. Huang J. Detecting neuronal currents with MRI: a human study. *Magn Reson Med*. 2014; **71**:756–762.

20. Hagberg GE, Bianciardi M, Maraviglia B. Challenges for detection of neuronal currents by MRI. *Magn Reson Imaging*. 2006; **24**:483–493.

21. Jiang X, Sheng J, Li H, et al. Detection of subnanotesla oscillatory magnetic fields using MRI. *Magn Reson Med*. 2016; **75**:519–526.

22. Qian Y, Lin YC, Hejazi S, et al. Quantum Sensing MRI for Noninvasive Detection of Neuronal Electrical Activity in Human Brains. *arXiv*. January 2026; 2601.16423.

23. Degen CL, Reinhard F, Cappellaro P. Quantum sensing. *Rev Mod Phys*. 2017; **89**(3):035002.

24. Ogawa, S. et al. Brain magnetic resonance imaging with contrast dependent on blood oxygenation. *Proc. Natl. Acad. Sci. U.S.A*. **87**, 9868–9872 (1990).

25. Kwong, K. K. et al. Dynamic magnetic resonance imaging of human brain activity during primary sensory stimulation. *Proc. Natl. Acad. Sci. U.S.A.* **89**, 5675–5697 (1992).